# Features of the study of binder state near the filler particles in elastomeric composites using an atomic-force microscope

O. K. Garishin[a], R. I. Izyumov[a,b], A. L. Svistkov[a,b]

[a] Institute of Continuous Media Mechanics Ural Branch of Russian Academy of Science
Perm, Korolev St., 1, 614013, Perm
email: gar@icmm.ru

[b] Perm State National Research University, Bukireva St. 15, 614990, Perm
email: email: izumov@icmm.ru

Atomic force microscopy (AFM) is one of the most promising methods for investigating the structure of materials at the micro and nanoscale levels, as well as their local physical-mechanical properties (which, as already known, can be very different from macroproperties). The experimental data obtained with the help of AFM are not very informative in themselves. Therefore, their further theoretical interpretation with the use of various mathematical and physical models is required. In addition, there is also a serious problem of assessing the accuracy and reliability of the results obtained for AFM. This is especially pronounced when studying materials with a strong mechanical structural heterogeneity. Such materials include considered in this paper elastomers, filled with rigid dispersed particles.

The article presents a technique for mathematical processing of such experimental data, which allows to significantly improve the accuracy of their interpretation. It is based on the use of three criteria for filtering and processing the initially obtained experimental information.

1) The criterion of the "relief of the nanoscale", which makes it possible to extract a relief with objects of low curvature from the overall picture obtained. It is on it that the filler particles protruding on the surface are clearly visible (usually covered with a thin film of elastomeric binder, known in the literature as bound rubber).

2) "Adhesion deflection" criterion, by means of which it is possible to isolate the filler particles by changing the adhesion forces between the AFM probe and the sample surface.

3) The criterion of "indentational compliance", designed to determine the location and stiffness of the filler particles.

All these criteria were used to decode the AFM scanning data of the surface of disperse filled elastomeric composite. Optimal conditions for the application of each of the criteria were determined from the analysis of the results obtained.

Also, the accuracy of the experimental results was estimated on the basis of a comparison of the data obtained with the forward and reverse horizontal motion of the AFM probe when scanning the same surface area.

## 1. Introduction.

One of the priority areas of modern materials science is the creation of new nanostructured materials with improved operational properties (these are the so-called nanocomposites). The solution of this problem is impossible without deep fundamental knowledge of the internal structure of such materials and their local physicomechanical properties at the micro, meso, and nanoscale levels. Effective management of the processes and phenomena occurring at these scales makes it possible to purposefully create materials with fundamentally new characteristics unattainable with the use of traditional technologies. Nowadays atomic force microscopy (AFM) is one of the most promising tools for such studies [1-5]. Its main advantage over conventional electron microscopy is that the AFM allows us to obtain information not only about the topology of the internal structure of the material, but also about its local physical properties. In 1986 Gerd Bining, Calvin Quite and Christopher Gerber [6] received the Nobel Prize for the creation of the invention of a scanning atomic force microscope. At present, their invention is widely used in various fields of modern science - physics, chemistry, biology, etc. AFM is also successfully used in materials science in the study of morphology and local physical and mechanical properties of the material at the nanostructural level (on scales where effects related to the molecular structure of the material must already be taken into account, although the material itself can still be considered as a continuous medium). Today it is already well known that the physical properties of nanoparticles and clusters, determined by their extremely high specific surface, can differ very much from the macroscopic characteristics [7]. With the help of AFM, scientists can now determine nanostructural local elastic moduli [8-15], hardening parameters [16, 17], creep [18], residual stresses [19]. These technologies allow one to directly observe and quantify such nanoscale phenomena as the appearance of dislocations, the appearance of shear instability, phase transitions and many other phenomena that are inaccessible to previously known technologies [20-24].

The principle of AFM is based on the force interaction between the studied surface and the cantilever beam (cantilever) with a sharp silicon probe at the free end. Usually this probe has the shape of a cone with a rounded apex. The length of the beam is about 100-200 μm, the height of the cone is 1-3 μm. The radius of the tip of the probe (which determines the resolution of the device) in modern cantilevers varies from 1 to 50 nanometers. The force acting on the probe from the surface leads to the bending of the console. By recording the amount of bending, it is possible to control the force of interaction between the probe and the object under study. In modern microscopes, optical methods are used for this. The optical system of the AFM is aligned in such a way that the radiation from the semiconductor laser is focused on the probe bracket and the reflected beam hits the photosensitive sensor. A fairly complete popular description of the principles of AFM's work can be found, for example, in [25, 26].

Nanoindentation methods are especially successful in the science of polymers. Most polymers are often much softer than the material of the AFM probe, which allows a deep enough penetration inside the sample during indentation. As a result, studies of the probe penetration process can give unique information about the mechanical properties of the material at the nanostructural level.

During the experiment, the AFM probe scans the selected sample surface. The resulting data are the dependencies between the coordinates of the scan points ($x, y$) and the vertical coordinate of the probe base ($z$) and the deviation of the end of the cantilever $d$ from which the force acting on the probe $F$ [26] is calculated. These results themselves (without additional knowledge of the subject of research) are not very informative, so their further theoretical interpretation with the use of various mathematical and physical methods is required.

In addition, there is another very serious problem of evaluating the accuracy of the results obtained on the AFM. As far as we can judge from the analysis of known publications [27-33], in most papers the main attention is paid to the study of the influence of various factors (probe shape, probe effect on the sample, the effect of moisture, nonlinear piezoelectric effects) on the scan result. This is probably caused by the fact that conventional methods of control are inapplicable in this case. New specific techniques are required. One such approach with respect to dispersed-filled elastomers is the subject of this article.

2. **Criteria for determining the surface areas of a sample in which the filler particles are located near the boundary of the material with air**

   2.1. **Criterion for the allocation of surface areas using the relief of the nanoscale**

The surface of the samples studied by AFM has a complex relief. It can be represented as the sum of two reliefs [34]. It is a relief of small curvature (the surface of the macrolevel of the material) and a relief of high curvature (the surface of the nanoscale of the material). The last one shows the protruding filler particles, usually covered with a thin film of elastomeric binder, known in the literature as a bound rubber. The topography of the surface at the nanoscale can be distinguished by means of an integral filter determined by the following transformation:

$$z_{nano}(x, y) = \int_S \varphi(r) z(x_1, y_1) dS,$$

where $z_{nano}$ is the height of the point location above the horizontal surface; $z$ is the coordinate along the vertical axis at the current point of the integration region $S$; $x, y$ are the coordinates of the point of nanoscale on the horizontal plane; $x_1, y_1$ are the coordinates of the current point in the integration region S on the horizontal plane. The kernel of the integral operator has the form:

$$\varphi(r) = \delta(r) - C H(L-r),$$

$$r = \sqrt{(x-x_1)^2 + (y-y_1)^2},$$

where $\delta(\cdot)$ is the Dirac function; $H(\cdot)$ is the Heaviside function; $L$ is the characteristic size used to extract the topography of the nanoscale. The constant $C$ is determined from the normalization condition

$$\int_S \varphi(r) \, dS = 1.$$

Further we will use the experimental data obtained on the quadrate region of the sample surface, whose side is 1.5 μm (Fig. 1a). However, the side of the square area used for the study will be slightly smaller. This is caused by the need to exclude from consideration points located at a distance $L$ from the boundary and closer, which is necessary for the application of an integral filter in order to obtain a nanoscale relief. In addition, the size of the region will be reduced in accordance with the need to quantify the accuracy of the experimental data. This will be discussed later. As a result of using the filter, the relief shown in Fig. 1b is obtained. For calculations we chose the characteristic size $L = 25$ nm.

The next step is to get rid of noise near the zero coordinate on the relief of the nanoscale surface and highlight the vertices on this relief. To achieve this goal, we cut off parts of the relief with a height below $C_z = 0.03L$ (Fig. 1c). The peaks above this plane determine the areas where the filler particles are expected to be near the boundary of the material. Further the determination of the probable location of particles on the basis of nanoscale surface analysis will be called "the first criterion". It is formulated by the following inequality:

$$z_{nano} < C_z$$

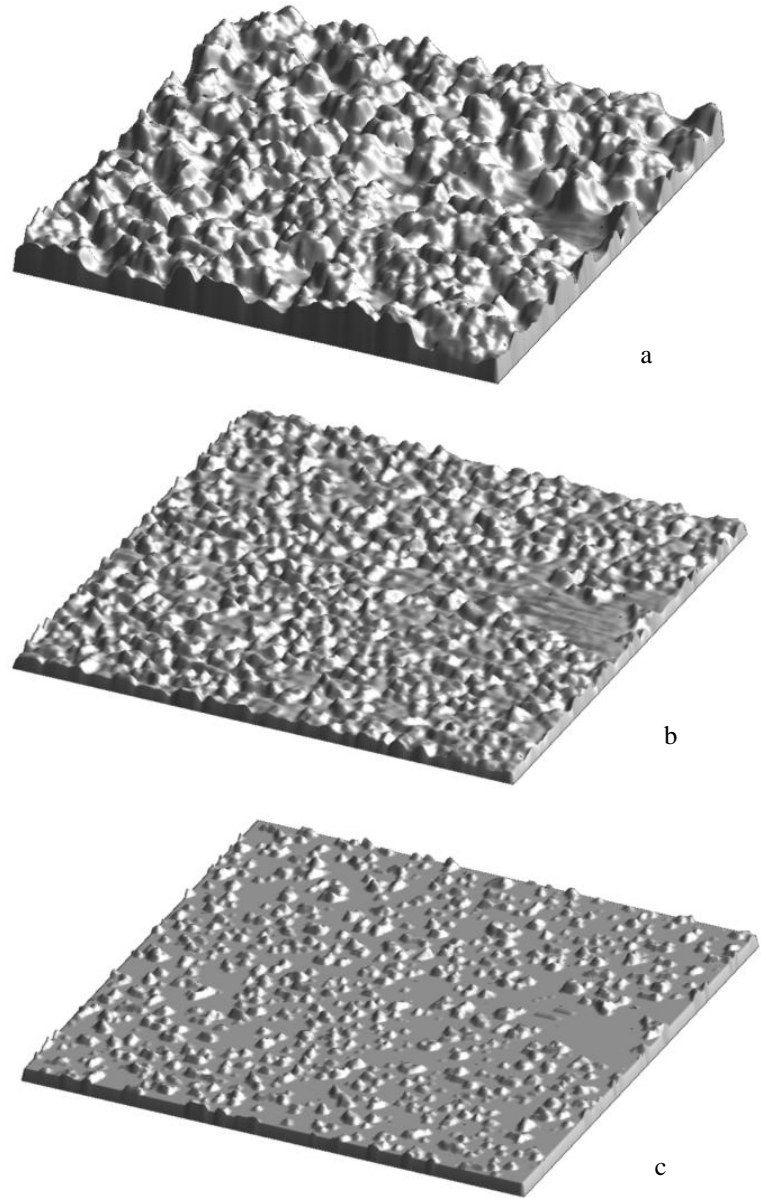

a

b

c

Figure 1. Relief of the studied sample surface: (a) – before conversion with the help of an integral filter and after: (b) – the surface of high curvature (nanoscale surface) and (c) – the vertices of the nanoscale surface.

## 2.2. Criteria for the allocation of surface areas using the parameters of the material's indentation curve

In the literature, the parameters shown in Figure 2 are used to analyze information of the material mechanical properties obtained by the nanoindentation modes of scanning. This parameters are the displacement of the cantilever base $\Delta z$, the depth of probe penetration into the material $u$, the deviation of the probe relative to the base of the cantilever $d$ (which is counted from the unloaded state), the distance from the current position of the probe to the point of its greatest penetration into the material $\gamma$ (which in the english-language literature is called "separation"). The direction of the counting of the parameters from the zero value is indicated by the arrows in Fig. 2.

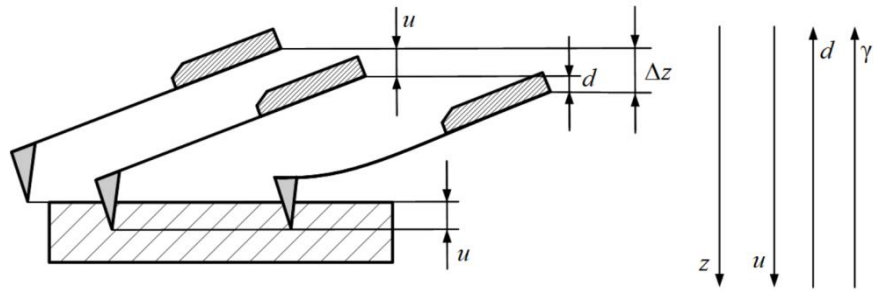

Figure 2. The illustration showing the physical meaning of the nanoindentation parameters and the direction of the axes for reading the corresponding parameters

The second criterion for finding the surface regions in which the filler particles are located near the material boundary is the minimum deviation of the tip of the probe $d_A$, taken with the opposite sign (Fig. 3). The minimum deviation is determined on the reverse course of the probe, when it is extracted from the material. This is the moment when the elastomer that adheres to the probe begins to detach or slip from the probe

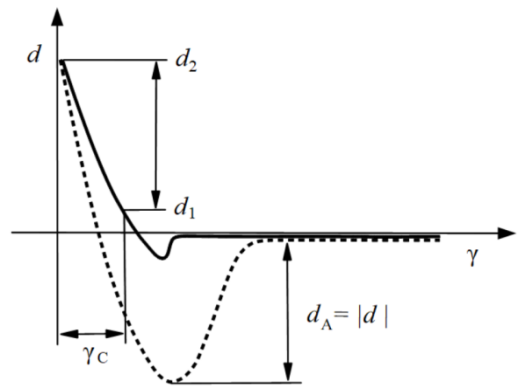

Figure 3. Nanoindentation curve determining the dependence of the probe deviation relative to the cantilever base on the parameter of probe removal from the point of its greatest penetration into the material

Further, the parameter $d_A$ will be called "the adhesion deviation". Why does it allow us to estimate the presence of filler particles near the surface of the sample? We consider elastomeric materials with active nanofillers. It has high interaction energy with polymer molecules. This means that if a particle is located below the surface, then when the probe is indentated into this region of the material, this particle will demonstrate significant resistance to penetration. Thus, the probe penetrates into the material to a shallower depth, so a smaller amount of elastomer adheres to it, and the less force is required to detach the probe from the polymer when the probe is retracted from the sample (Fig. 4).

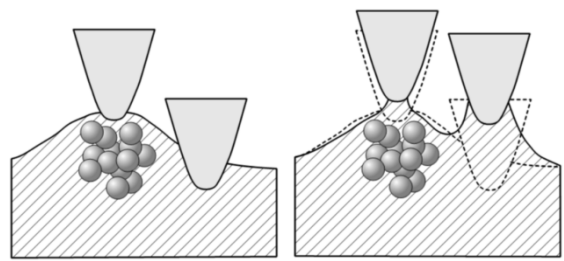

Figure 4. The illustration of the dependence of adhesion deviation $d_A$ on the maximum penetration of the probe into the material. The state (left) corresponds to the maximum penetration of the probe; (right) – the moment of detachment of the elastomer from the probe on the reverse course

Therefore, as the second criterion for the definition of filler particles, it is expedient to use condition:
$$d_A < C_A$$
The value of the $C_A$ parameter will be chosen from the condition that the area of projections of the selected sections on the horizontal plane in accordance with the second criterion coincides with the area of the projections of the sections selected according to the first criterion.

The third criterion we call "indentational compliance" $\gamma_c$. In the english literature it is called the "deformation" parameter. It is defined as follows. The maximum deviation of the probe relative to the cantilever base is denoted by the parameter $d_2$ (Figure 5.3). The minimum deviation at this stage of indentation is zero, but often there are pronounced features of the behavior of the probe (due to the influence of surface effects), which makes it difficult to accurately determine the zero deviation. In order to remove the unpleasant zone of the curve from consideration, an additional parameter $d_1$ is used, the value of which was chosen from the condition $d_1 = 0.15 \cdot d_2$. Then the difference between the corresponding values of the parameter $\gamma$ in the forward course (immersion into the material) is the indentational compliance $\gamma_c = (d_1) - (d_2)$. In this range of probe displacement, the specific feature of the interaction of the probe with the sample is that the main work is used to deform the material. In contrast to this, at a lesser depth of indentation, the other phenomena can predominate. The force acting on the probe is significantly affected by its interaction with the surface of the sample (the expenditure of energy going to form a new contact surface, a decrease in the area of contact between the sample and the probe with air, etc.).

An important point, which should be paid attention, is the case when the probe bounces off from the surface of the sample. It is connected with the following. We are investigating material that has both very hard areas and very soft. Therefore, it is impossible to adjust the atomic force microscope in such a way that the obtained experimental information on the entire surface has a satisfactory accuracy. The probe penetrates the material at a sufficiently high speed. Therefore, it is possible to obtain a probe bounce. For such points, the indentational compliance $\gamma_c$ has a negative value.

If a particle of filler is located under a thin layer of elastomer, then indentation of the probe will require considerable force. Therefore, the third criterion for the allocation of filler particles will use condition:
$$\gamma_c < C\gamma,$$
where the value of the parameter $C\gamma$ will be chosen from the condition that the area of the projections of the selected particles to the horizontal plane in accordance with the second criterion coincides with the area of the projections of the particles selected according to the first criterion.

**2.3. The accuracy of the experimental data obtained in the mode of mechanical properties mapping**

Estimation of the accuracy of experimental data plays an important role in their obtaining and as a result of their processing. First of all you should pay attention to the surface relief maps obtained in the trace and retrace modes. First of all, it is necessary to pay attention to the surface relief maps obtained at the "trace" and "retrace" scanning stages. They differ from each other. The "trace" stage is the movement of the cantilever along the surface of the material in one direction. "Retrace" is the movement in the opposite (return) direction.

To achieve the best match, you need to shift the obtained scans (data maps) relative to each other. Probably, such a disagreement is caused by the fact that the change in the direction of the probe movement takes some time, during which the information on several points is taken while moving in one direction, and is not taken into account when moving in the opposite direction. In the further study, the maps with the parameters of adhesive deflection and indentational compliance are shifted in the same way as the surface relief maps of the material.

The analysis of scans showed that there is still a noticeable difference in the surface relief obtained in the trace and retrace stages of scanning. Figure 5 compares the areas selected using the first criterion in the trace and retrace stages. But this difference is not so significant. The discrepancy occupies about 6 percent of the area of the selected surface area.

Significantly worse results are obtained if we compare areas that satisfy different criteria. Figure 6 shows the comparison of maps of arithmetic average values of the corresponding parameters obtained at the trace and retrace stages. These parameters are: the height of the relief of the nanoscale, the adhesion deflection, the indentational compliance. The difference in this case is 10 or more percent.

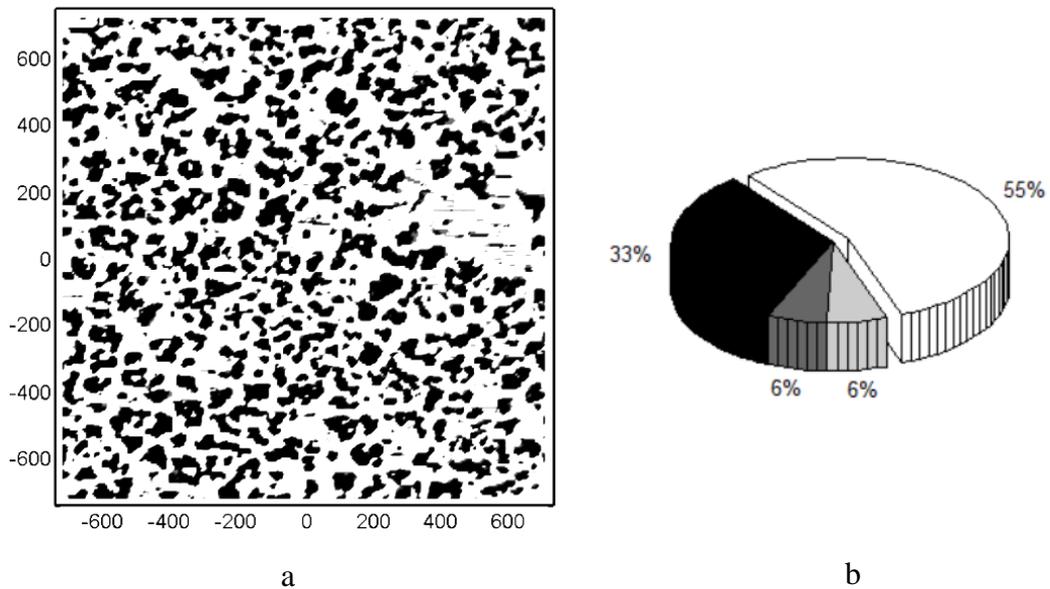

Figure 5. (a) Highlighted areas, where particles of the filler are located shallow beneath the surface, (b) Surface fraction diagram: black color – areas that satisfy the criterion for map of heights obtained in both trace and retrace stages; dark gray color – only in trace stage; light gray color – only in retrace stage.

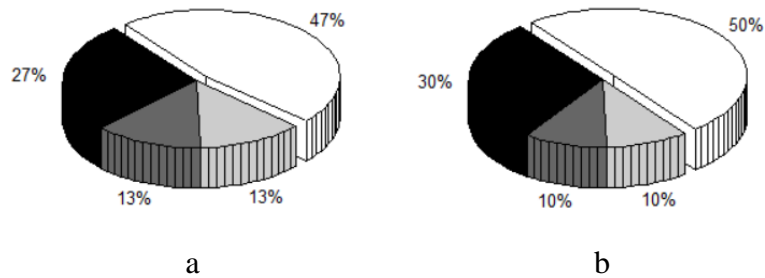

Figure 6. Surface fraction diagram: (a) area that satisfy the criterion for map of heights and ndentational compliance; (b) area that satisfy the criterion for map of heights and adhesive deviation. Black color corresponds to the simultaneous satisfaction of two criteria; dark gray – only the criterion for heights; light gray – only the second criterion.

However, we can pay attention to the following fact. There are areas where all three criteria are simultaneously satisfied (Fig. 7a). This allows us with sufficient confidence to assume that the presence of filler particles is guaranteed under the surface of these regions. Moreover, the experimental data allow us to note the possible feature of the interaction of the probe with the material. These are situations where the probe hits the surface of the sample and bounces off from it (Fig. 7b) (and value of the indentational compliance is negative). The interpretation of the interaction of the probe with the material requires consideration of dynamic effects, the accuracy of which is difficult to determine. Therefore, in our opinion, it is advisable to limit the analysis of indentation curves by regions where three criteria are satisfied simultaneously, and in which there is no bounce of the probe from the surface.

In the example (Fig. 7), for highlighted areas the average values of the adhesion deflection and the indentational compliance are 10.3 nm and 1.5 nm, respectively. The root-mean-square deviation of the adhesive deflection and adhesion compliance values is 2.3 nm and 1.1 nm. Further interpretation of these quantities requires the use of mathematical models of nanoindentation.

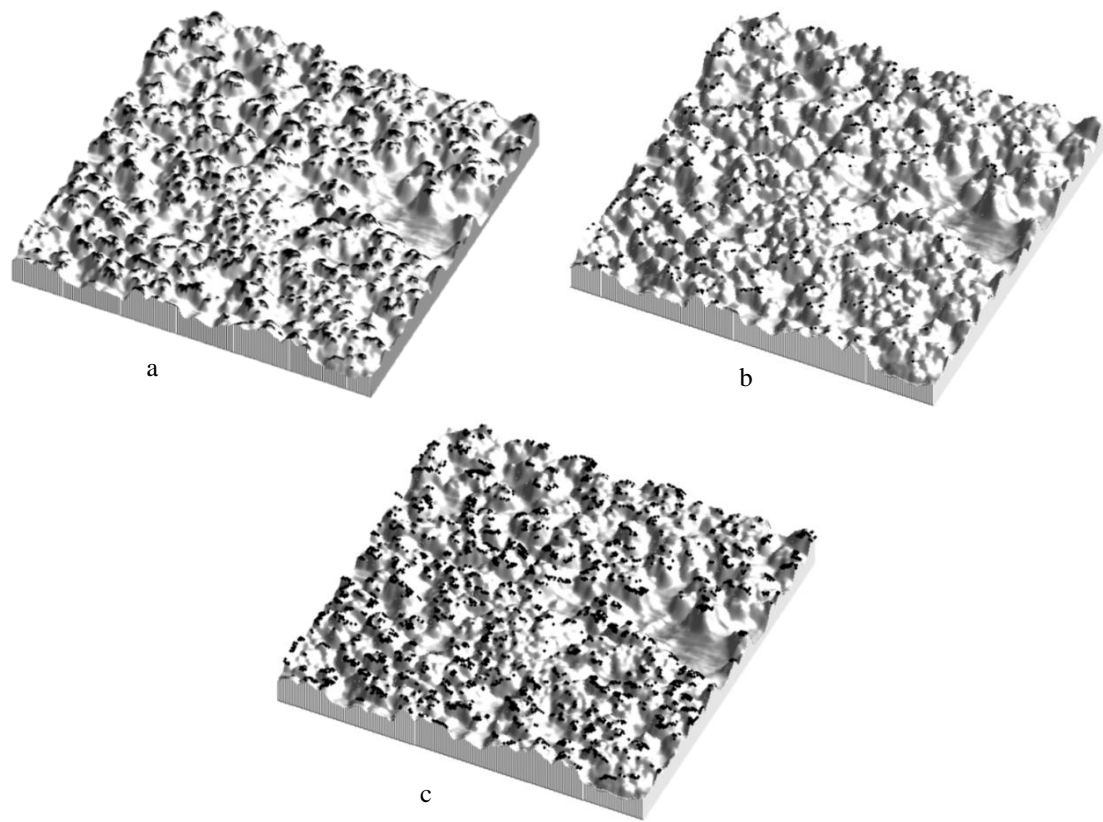

Figure 7. The relief of the investigated area. Black color: (a) – three criteria for the location of the filler particles near the material surface are satisfied simultaneously; (b) – there was a bounce of the probe from the surface; (c) - three criteria are satisfied, and there is no bounce of the probe.

3. **Conclusion**

The performed researches have shown that an estimation of the accuracy of the results is an indispensable condition for the processing of AFM-data of the structure and properties of elastomeric nanocomposites. It is proposed to perform such an evaluation by comparing the maps of parameters obtained at the trace and retrace stages of scanning. The selection of areas of material where the filler particles are located near the surface is carried out using the criteria "nanoscale topography", "adhesion deflection" and "indentational compliance". It is established that criteria application makes it possible to determine the difference on surface areas. It is possible to select such areas of the surface where all three criteria are satisfied simultaneously. These areas can be used to determine the possible thickness of the elastomer layer, the rigidity, the interaction force with the probe, and the accuracy of the investigated quantities.

**Acknowledgements.** This work was supported by the Russian Foundation for Basic Research (grant №16-08-00914). The authors also thank Morozov I.A. for the help and provided experimental data.